\documentclass[11pt, notitlepage]{article}

\usepackage{a4}

\usepackage{xcolor}
\usepackage{algorithm}
\usepackage[noend]{algpseudocode}

\definecolor{TUMblue}{RGB}{0, 101, 189}
\definecolor{TUMlightblue}{RGB}{100,160,200}
\definecolor{TUMgreen}{RGB}{162,173,0}
\definecolor{TUMorange}{RGB}{227,114,034}
\definecolor{TUMivory}{RGB}{218,215,203}

\usepackage{natbib}

\usepackage{hyperref}
\hypersetup{
colorlinks=true,
linkcolor=TUMblue,
citecolor=TUMblue,
filecolor=TUMblue,
urlcolor=TUMblue
}

\usepackage{etoolbox}
\makeatletter

\pretocmd{\NAT@citex}{%
\let\NAT@hyper@\NAT@hyper@citex
\def\NAT@postnote{#2}%
\setcounter{NAT@total@cites}{0}%
\setcounter{NAT@count@cites}{0}%
\forcsvlist{\stepcounter{NAT@total@cites}\@gobble}{#3}}{}{}
\newcounter{NAT@total@cites}
\newcounter{NAT@count@cites}
\def\NAT@postnote{}

\def\NAT@hyper@citex#1{%
\stepcounter{NAT@count@cites}%
\hyper@natlinkstart{\@citeb\@extra@b@citeb}#1%
\ifnumequal{\value{NAT@count@cites}}{\value{NAT@total@cites}}
{\ifNAT@swa\else\if*\NAT@postnote*\else%
\NAT@cmt\NAT@postnote\global\def\NAT@postnote{}\fi\fi}{}%
\ifNAT@swa\else\if\relax\NAT@date\relax
\else\NAT@@close\global\let\NAT@nm\@empty\fi\fi
\hyper@natlinkend}
\renewcommand\hyper@natlinkbreak[2]{#1}

\patchcmd{\NAT@citex}
{\ifNAT@swa\else\if*#2*\else\NAT@cmt#2\fi
\if\relax\NAT@date\relax\else\NAT@@close\fi\fi}{}{}{}
\patchcmd{\NAT@citex}
{\if\relax\NAT@date\relax\NAT@def@citea\else\NAT@def@citea@close\fi}
{\if\relax\NAT@date\relax\NAT@def@citea\else\NAT@def@citea@space\fi}{}{}

\makeatother

\usepackage{amssymb, graphicx}
\usepackage{subfigure}
\usepackage{amsmath}
\usepackage{amsthm}
\usepackage{undertilde}
\usepackage{verbatim}
\usepackage{bbm}
\usepackage{ dsfont }
\usepackage{geometry}
\usepackage{pdflscape}
\usepackage{multirow}
\usepackage{aliascnt}

\newcommand{\mynewtheorem}[2]{
\newaliascnt{#1}{dummy}
\newtheorem{#1}[#1]{#2}
\aliascntresetthe{#1}
\expandafter\def\csname #1autorefname\endcsname{#2}
}

\theoremstyle{definition}
\mynewtheorem{thm}{Theorem}
\mynewtheorem{defi}{Definition}
\mynewtheorem{lem}{Lemma}
\mynewtheorem{cor}{Corollary}
\mynewtheorem{prop}{Proposition}
\mynewtheorem{exa}{Example}
\mynewtheorem{alg}{Algorithm}
\mynewtheorem{rem}{Remark}
\mynewtheorem{bsp}{Example}

\def\equationautorefname~#1\null{Equation~(#1)\null}

\newcommand{\aref}[1]{\hyperref[#1]{Appendix~\ref{#1}}}

\usepackage{sectsty}
\allsectionsfont{\sffamily}

\usepackage{url}

\usepackage{caption}
\usepackage{footnote}

\newcommand{\Xb}{\mathbf{X}}

\newcommand{\be}{\begin{equation}}
\newcommand{\ee}{\end{equation}}

\DeclareMathOperator{\dist}{dist}

\begin{document}
	
		\title{\textbf{\sffamily Stress Testing German Industry Sectors:\\ Results from a Vine Copula Based Quantile Regression		 
			}}
			
			\date{\small \today}
			\author{Matthias Fischer\footnote{Lehrstuhl f\"ur Statistik und \"Okonometrie, Lange Gasse 20, 90403 N\"urnberg; \href{mailto:matthias.fischer@fau.de}{matthias.fischer@fau.de}},\hspace{.2cm} Daniel Kraus\footnote{Zentrum Mathematik, Technische Universit\"at M\"unchen, Boltzmanstra\ss e 3, 85748 Garching; \href{mailto:daniel.kraus@tum.de}{daniel.kraus@tum.de}},\hspace{.2cm} Marius Pfeuffer\footnote{Corresponding author, Lehrstuhl f\"ur Statistik und \"Okonometrie, Lange Gasse 20, 90403 N\"urnberg; \href{mailto:marius.pfeuffer@fau.de}{marius.pfeuffer@fau.de}}\hspace{.2cm} and Claudia Czado\footnote{Zentrum Mathematik, Technische Universit\"at M\"unchen, Boltzmanstra\ss e 3, 85748 Garching; \href{mailto:cczado@ma.tum.de}{cczado@ma.tum.de}}}

			\maketitle
			
		

	\begin{abstract}
Measuring interdependence between probabilities of default (PDs) in different industry sectors of an economy plays a crucial role in financial stress testing. Thereby, regression approaches may be employed to model the impact of stressed industry sectors as covariates on other response sectors. We identify vine copula based quantile regression as an eligible tool for conducting such stress tests as this method has good robustness properties, takes into account potential nonlinearities of conditional quantile functions and ensures that no quantile crossing effects occur. We illustrate its performance by a data set of sector specific PDs for the German economy. Empirical results are provided for a rough and a fine-grained industry sector classification scheme. Amongst others, we confirm that a stressed automobile industry has a severe impact on the German economy as a whole at different quantile levels whereas e.g., for a stressed financial sector the impact is rather moderate. Moreover, the vine copula based quantile regression approach is benchmarked against both classical linear quantile regression and expectile regression in order to illustrate its methodological effectiveness in the scenarios evaluated.

	\end{abstract}
	\noindent \textit{Keywords:} stress testing; quantile regression; vine copulas; expectile regression








\section{Motivation}
Generally speaking, stress testing identifies potential vulnerabilities of financial institutions under hypothetical or historical scenarios. Financial institutions typically perform stress tests to assess possible short-term losses resulting from various types of risk (e.g., credit risk, market risk, operational risk).
The history of stress tests in the banking industry dates back to the early 1990s, where large banks started to initiate internal stress exercises. In 1996, the Basel Capital Accord was amended which requires banks to conduct stress tests and determine their ability to respond to market events. However, up until 2007, stress tests were typically performed only by the banks themselves, for internal self-assessment. Beginning in 2007, regulatory institutions became interested in conducting their own stress tests to ensure the effective operation of financial institutions. Since then, stress tests have been routinely performed by financial regulators in different countries or regions, to ensure that the banks under their authority are engaging in practices likely to avoid negative outcomes. Recently, the European Banking Authority (EBA) published the results of the 2016 EU-wide stress test of 51 banks. The aim of this stress test was to assess the resilience of EU banks to adverse economic developments. Similarly, the Federal Reserve Board currently published the scenarios to be used by banks and supervisors for the 2017 Comprehensive Capital Analysis and Review (CCAR) and Dodd-Frank Act stress test exercises.
The focus within this work is on stress testing credit risk and losses associated therewith. Typically, the loans portfolios of the banking sector consist of financial obligations to counterparties from different business lines. Hence, the key source of credit risk in that portfolio is that a counterparty may default, which would result in losses for the bank. Secondly, one has to be aware of (industry/country) sector risk which means that multiple clients within an industry sector default because of structural weaknesses within this sector. Finally, even counterparties from different sectors may tend to default together in economic downturns, e.g., if only one (industry/country) sector enters a crisis and these sectors are highly correlated with the other. In all cases, the primary interest lies in the explanation (in a first step) and the initiation of stress (in a second step) of the counterparties’ probability of default (PD).

In this context, quantile regression (QR) is an increasingly important empirical tool in economics and other sciences for analyzing the impact a set of regressors (e.g., macro-economic variables) has on the conditional distribution of an outcome (here:\ PD). Extremal QR, or QR applied to the tails, is also very helpful in the context of stress testing. The quantile regression method is used to estimate parameters in accordance with the distribution of a dependent variable. One could use high quantile values of these distributions such as the 95th percentile value for a period in which large stresses occur. For a period in which no stresses are exerted on the economy, normal parameters estimated by the least square method might come to application.

Extremal QR was applied by \cite{koenker2002inference}, \cite{schechtman2012macro}, \cite{covas2014stress} and \cite{ong2014guide} within a macro-credit risk link context, i.e., connecting PDs and macro-economic variables. In contrast to these studies, we apply extremal QR to investigate how crises (in the sense of higher PDs) in industry sectors are connected to other industry sectors. Above that, we circumvent the disadvantages of classical QR, namely the occurrence of so-called quantile-crossings and apply D-vine copula based quantile regression as recently advocated by \cite{kraus2017d}, instead.
Against this background, the outline of this work is as follows: \autoref{sec:Dvines} briefly reviews D-vines and D-vine copula based regression. \autoref{sec:data} is dedicated to the empirical study which relies on (averaged) default probabilities of 9 different German industry sectors and selected subgroups. After a brief description of the underlying data set, the data transformation to the copula scale and the resulting data structure is discussed. A D-vine based copula regression is performed and highlights are presented. Finally, we illustrate its superiority to alternative approaches such as traditional quantile regression and expectile regression. \autoref{sec:conclusion} concludes.

\section{A short review on D-vines and D-vine copula based quantile regression}\label{sec:Dvines}
Copulas are important and useful tools to model the dependence between financial variables. They allow separate considerations of marginal distributions and dependencies due to Sklar's Theorem \citep{Sklar}, which states that the joint distribution $F$ of a continuous random vector $\Xb=(X_1,\ldots,X_d)$ can be expressed in terms of its marginal distributions $F_j$, $j=1,\ldots,d$, and its unique copula $C:[0,1]^d\rightarrow [0,1]$ through the relationship
\[
F(x_1,\ldots,x_d)=C(F_1(x_1),\ldots,F_d(x_d)).
\]
This facilitates flexible modeling of multivariate distributions, since the marginals and the dependence function can be modeled separately. Another consequence of Sklar's Theorem is that, as long as the interest lies in the dependence between random variables, one can consider the probability integral transformed random variables $U_j=F_j(X_j)$, $j=1,\ldots,d$, which are uniformly distributed. We say that these variables and their realizations are on the copula scale. Thorough introductions to copulas are given in \cite{joe1997multivariate} and \cite{nelsen2006introduction}.\\
In high dimensions many parametric copula models lack flexibility. For example, the exchangeable Archimedean copulas only have one or two dependence parameters and elliptical copulas always exhibit symmetric dependencies in the tails. Vine copulas overcome these shortcomings by breaking down the modeling of a multivariate copula to the fitting of several bivariate copulas \citep{bedford2002vines,aasczado,fischeretal}. E.g., in three dimensions a vine copula density decomposition is given by
\begin{equation}\label{eq:vine3d}
c(u_1,u_2,u_3)=c_{12}(u_1,u_2)\,c_{23}(u_2,u_3)c_{13;2}\left(C_{1|2}(u_1|u_2),C_{3|2}(u_3|u_2);u_2\right),
\end{equation}
where the three pair-copulas $c_{12}$, $c_{23}$ and $c_{13;2}$ can be modeled separately, each with its own parametric copula family and dependence parameter(s). The arguments of the conditional copula $c_{13;2}$, namely $C_{1|2}(u_1|u_2)$ and $C_{3|2}(u_3|u_2)$, are obtained by taking derivatives, i.e., $C_{j|2}(u_j|u_2)=\partial/\partial u_j C_{j2}(u_j,u_2)$, $j=1,3$. Note that choosing only bivariate Gaussian copulas in the decomposition always results in a multivariate Gaussian copula. Thus, vine copulas can be seen as a generalization of Gaussian copulas.\\
To conduct statistical inference it is usually assumed that the conditional copula $c_{13;2}$ does not depend on $u_2$, i.e., $c_{13;2}(\cdot,\cdot;u_2)\equiv c_{13;2}(\cdot,\cdot)$. This so-called simplifying assumption has been inspected by many researchers \citep[see e.g.,][]{stoeber2013simplified,spanhel2015simplified,killiches2016examination}. While there exist cases where it is a model restriction, empirical studies show that the assumption is not a severe restriction for financial data sets \citep{kraus2017growing}.\\
A special subclass of vine copulas that we will use to model the dependence between default probabilities of German industry sectors are D-vine copulas. The decomposition of a $d$-dimensional simplified D-vine copula density can be seen as a generalization of \autoref{eq:vine3d} and is given by
\begin{multline}
c(u_1,\ldots,u_d) = \prod_{i=1}^{d-1}\prod_{j=i+1}^dc_{ij;i+1,\ldots,j-1}\big(C_{i|i+1,\ldots,j-1}\left(u_i|u_{i+1},\ldots,u_{j-1}\right),\\
C_{j|i+1,\ldots,j-1}\left(u_j|u_{i+1},\ldots,u_{j-1}\right)\big).
\label{eq:D-vine_density}
\end{multline}
Thus, modeling a $d$-dimensional D-vine copula consists of fitting $d(d-1)/2$ bivariate parametric copulas, so-called pair-copulas. The arguments of the pair-copulas are obtained from a recursive formula given in \cite{joe1997multivariate} using the specified pair-copulas.\\
For the purpose of stress testing we are interested in estimating conditional quantiles $q_{\alpha}$ of a response variable $V=F_Y(Y)$ conditioned on predictor variables $U_1=F_1(X_1),\ldots,U_d=F_d(X_d)$:
\begin{equation}
q_{\alpha}(u_1,\ldots,u_d):=C^{-1}_{V|U_1,\ldots,U_d}(\alpha|u_1,\ldots,u_d)
\label{eq:defi_q_alpha}
\end{equation}
Given a fully specified D-vine copula with response $V$ as first variable, the conditional copula quantile function $C^{-1}_{V|U_{1},\ldots, U_{d}}(\alpha|u_{1},\ldots,u_{d})$ can be analytically expressed only using the pair-copulas of the D-vine \citep[see][]{kraus2017d}.\\
The only question remaining is how to fit the D-vine to given copula data, such that predictions of its conditional quantile functions are optimal. \cite{kraus2017d} propose an algorithm, which sequentially adds covariates to the D-vine that improve the model fit (measured in terms of the log-likelihood of the conditional density $c_{V|U_1,\ldots,U_d}(v|u_1,\ldots,u_d)$) the most. This is done until none of the remaining covariates is able to improve the model fit. Thus, the algorithm facilitates an automatic forward covariate selection. The authors further demonstrate that D-vine copula based quantile regression outperforms traditional quantile regression methods established in the literature regarding prediction accuracy.\\

\section{Data description and empirical results}\label{sec:data}

\subsection{Original data set}

\begin{figure}[htbp]
	\centering
	\includegraphics[width=0.65\linewidth]{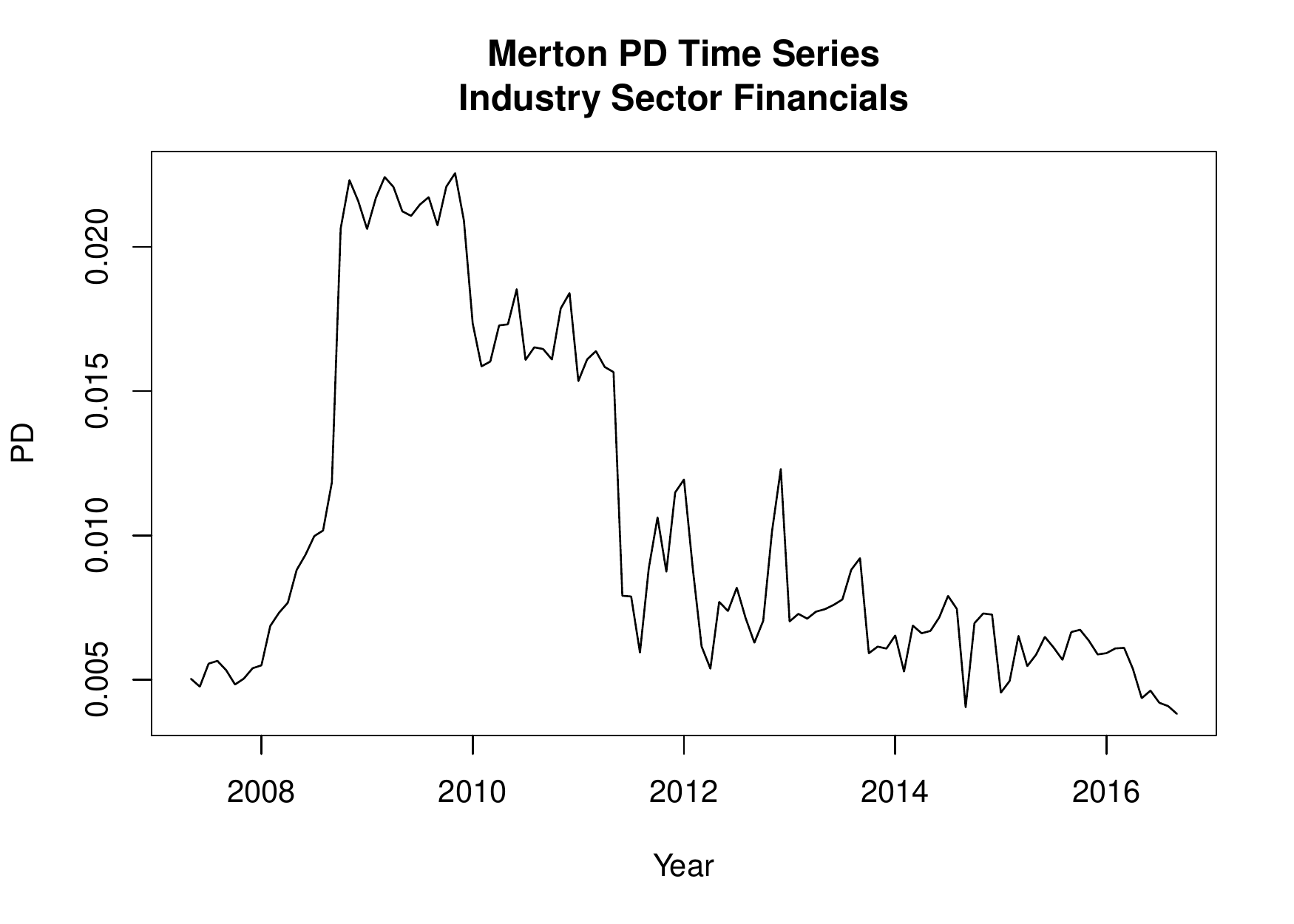}
	\caption{Exemplary plot of the mean probability of default in the industry sector Financials.}\label{fig:tsfin}
\end{figure}

For parameter estimation, a data pool with German exchange traded corporates is used. Based on a standard Merton model \citep[see][]{merton1974pricing}, the original time series consist of one-year PDs between May 2007 and September 2016. The PDs which are available monthly over the time horizon and on company level are averaged on sector level, using a market-based classification scheme, similar to the GICS and ICB systems with 9 industry sectors (Basic Materials, Communications, Cyclical Consumer Goods \& Services, Non-cyclical Consumer Goods \& Services, Energy, Financials, Industrials, Technology, Utilities) and several industry groups for a finer classification scheme. More precisely, PD’s within a Merton setting estimate the probability that a firm will default over a specified period of time (here:\ one year). As usually, ``default'' is defined as failure to make scheduled principal or interest payments. In the Merton setting, a firm defaults when the market value of its assets falls below its liabilities payable. Hence, the relevant drivers are the current market value of the firm, the level of the firm’s obligations and the vulnerability/sensitivity of the market value to large changes. Exemplarily, \autoref{fig:tsfin} illustrates the “PD history” for Financials with significant peaks caused by the last financial crisis from 2007 to 2008.

\subsection{Time dependencies and transformation to copula scale}

\begin{figure}[htbp]
	\centering
	\includegraphics[width=0.99\linewidth]{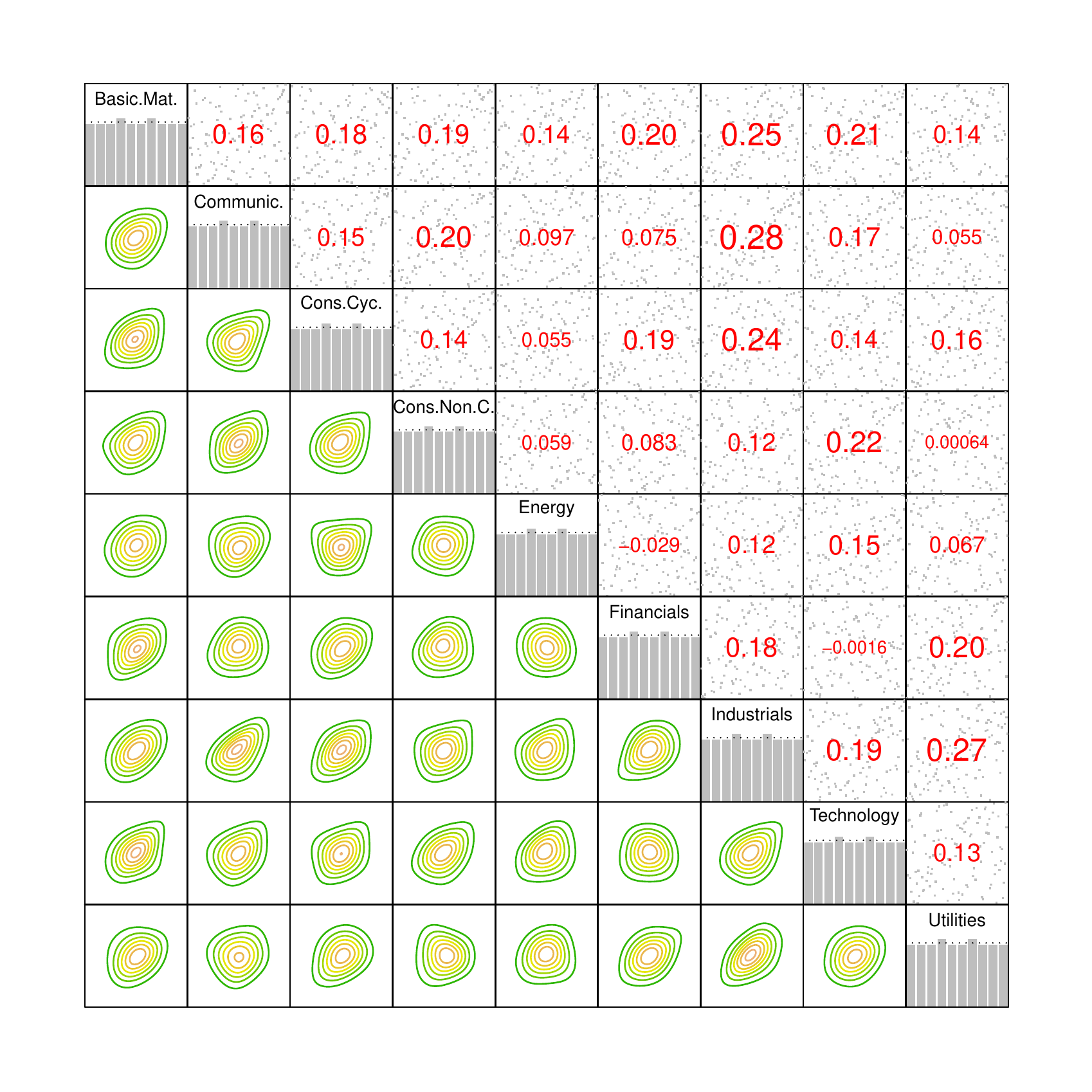}
	\caption{Upper triangular matrix: scatter plots and Kendall's $\tau$ values between pairs of aggregated sectors.\\
		Lower triangular matrix: empirical contour plots of copula densities of pairs of aggregated sectors.
		Diagonal: histograms of marginals.}
	\label{fig:pairs_plot}
\end{figure}

We consider the monthly differences of the aggregated sector PDs on a rough as well as on a more detailed level. The differenced data series are stationary and do not exhibit any autocorrelation or volatility clustering. Therefore, no ARMA-GARCH models are necessary to account for time dependencies.\\
Next, we transform the differenced (aggregated) data to the copula scale by applying the probability integral transform using the empirical cumulative distribution function as marginal distribution functions. The corresponding contour plots and Kendall's $\tau$ values are displayed in \autoref{fig:pairs_plot}.

The dependencies are weak to medium and mostly positive, at first glance. The Industrials sector seems to have the strongest interdependencies with the other ones. The empirical copula density contours suggest that the dependencies are quite asymmetric, such that Gaussian copulas or elliptical copulas in general would not provide reasonable fits. Some pairs seem to exhibit tail dependence (e.g., Industrials and Technology). The histograms of the marginals displayed on the diagonal naturally are flat after transforming the observations with their empirical distribution function.

\subsection{Selected results of the D-vine copula based quantile regression}

We perform stress tests similar to the ones described in \cite{kraus2017d}. Large values (i.e., close to 1) of the variables on the copula scale correspond to large differences in the sector PDs. Therefore, inducing stress on an industry sector will be treated as setting the value of the respective industry sector covariate to a predetermined quantile level $\kappa\in(0,1)$, usually $\kappa\in\{0.95,0.99\}$. Then we use D-vine quantile regression to examine the effect of the stressed companies (covariates) on the other companies (responses). The predicted quantile will give information on how strongly the response companies are affected by the stress scenario. E.g., large deviations of the conditional predicted mean from the unconditional median of 0.5 imply strong effects of the stress scenario.

\subsubsection{Results for stressing at 95\% and 99\% on aggregated data}
\label{sec:aggr}
\begin{figure}[htbp]
	\centering
	\includegraphics[width=0.49\linewidth]{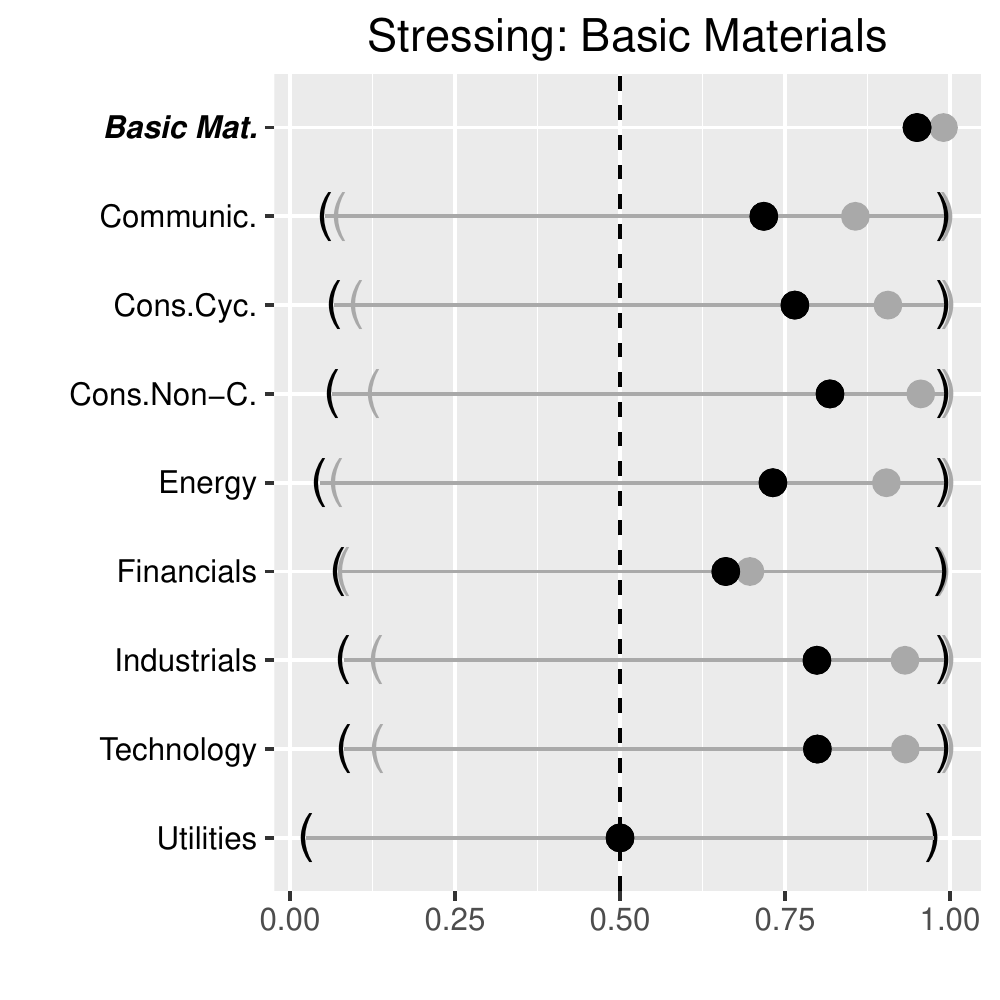}         
	\includegraphics[width=0.49\linewidth]{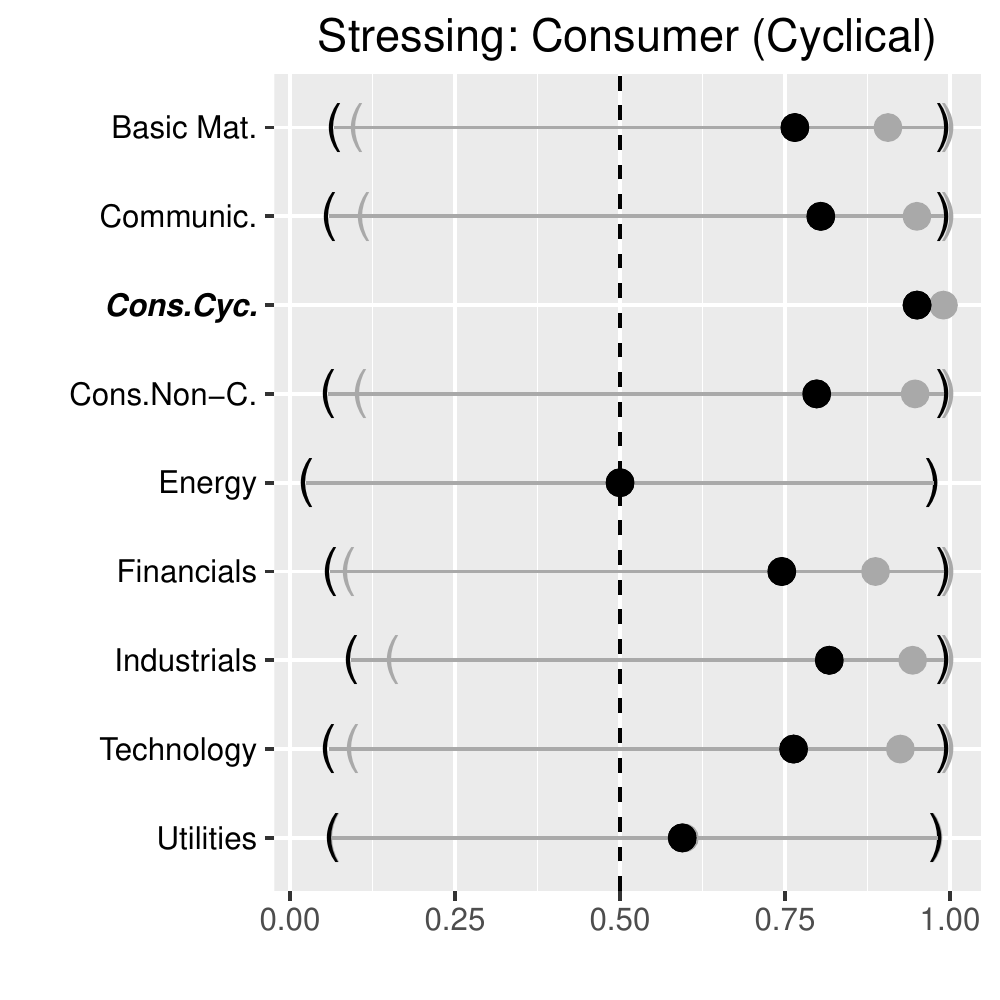}\\	       
	\includegraphics[width=0.49\linewidth]{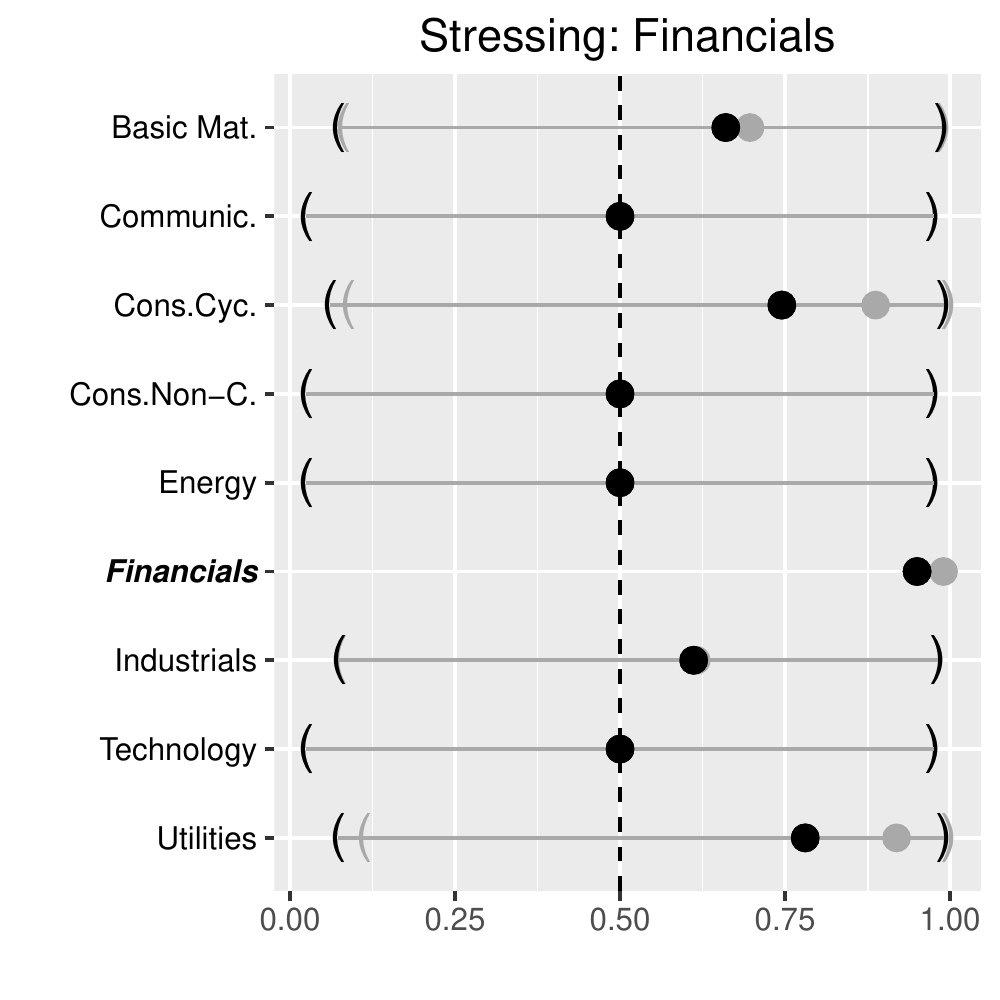}
	\includegraphics[width=0.49\linewidth]{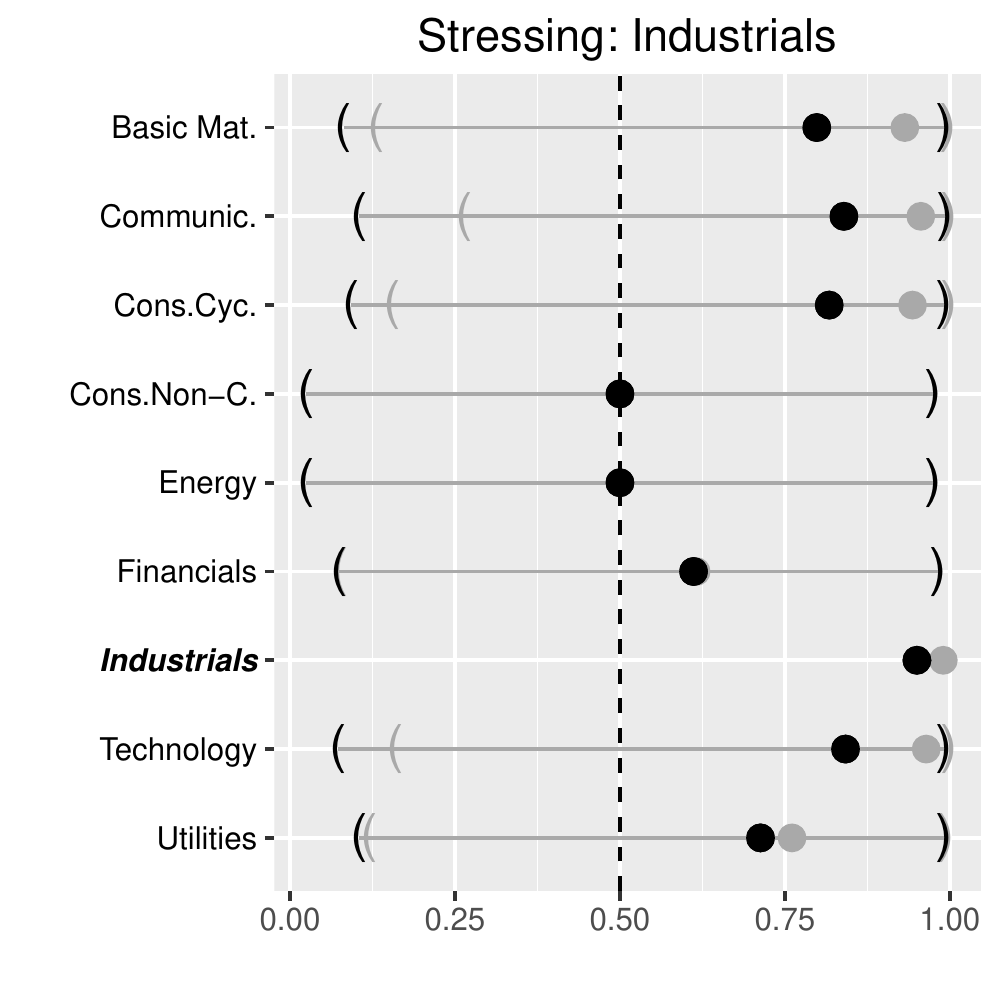}	
	\caption{Stress testing results for selected industry sectors. In each plot, the sector written in bold italics is stressed at levels 95\% (black) and 99\% (gray). The brackets indicate the 95\% prediction interval.}
	\label{fig:stressnew}
\end{figure}

At first we present the results on the aggregated data, stressing one sector at stress levels $\alpha = 0.95$ (black) and $\alpha= 0.99$ (gray). For reasons of brevity, our focus lies on the sectors Basic Materials (upper left panel of \autoref{fig:stressnew}), Cyclical Consumer Goods (upper right panel of \autoref{fig:stressnew}), Financials (lower left panel of \autoref{fig:stressnew}) and Industrials (lower right panel of \autoref{fig:stressnew}).

As expected, the effect on the conditional quantile functions strongly depends on the specific (stressed) sector and - to some minor extent - on the concrete stress level.

Across all sectors under consideration, Energy seems to be quite resistant against local sector crises. The same holds for the Utilities sector if we restrict ourselves to crises arising from Basic Materials and Cyclical Consumer Goods. On the other hand, sector crises arising from Basic Materials and Cyclical Consumer Goods spread over to most of the other sectors beside the Utilities sector. This does not hold for Financials and Industrials. In particular, stressing the sector Financials mainly affects the sector Cyclical Consumer Goods and Utilities. Above that, a simulated crisis in the Industrial sector has a significant impact on the segments Basic Materials, Communications, Cyclical Consumer Goods and Technology.

\subsubsection{Selected scenarios on detailed level}

Next, we focus on an industry classification scheme consisting of 55 sub-sectors in order to perform stress tests and analyze stress effects on a more granular scale. For instance, specific sub-sectors can now be isolated in order to check which of the other sub-sectors are affected and how strong these effects can be expected. However, it should be mentioned that the number of companies which are used to calculate averaged (sub-)sector PD’s decreases with increasing granularity which also implies decreasing statistical precision of the estimators. As before, we consider the probability integral transforms of the differenced time series data.

Again, we restrict ourselves to two arbitrarily selected industry sectors out of the four sectors discussed above: Basic Materials and the Automobile Industry.

\begin{figure}[htbp]
	\centering
	\includegraphics[width=0.49\linewidth]{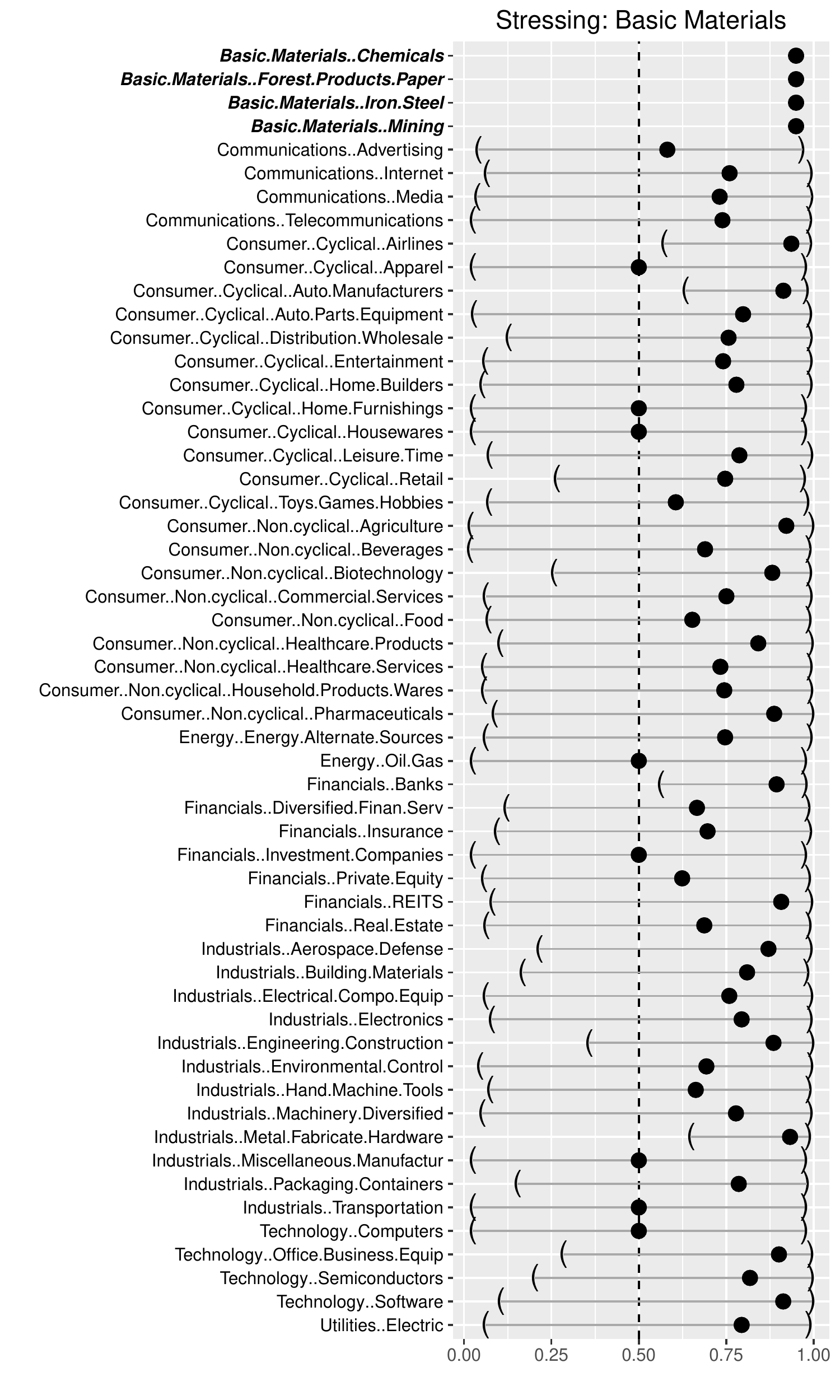}
	\includegraphics[width=0.49\linewidth]{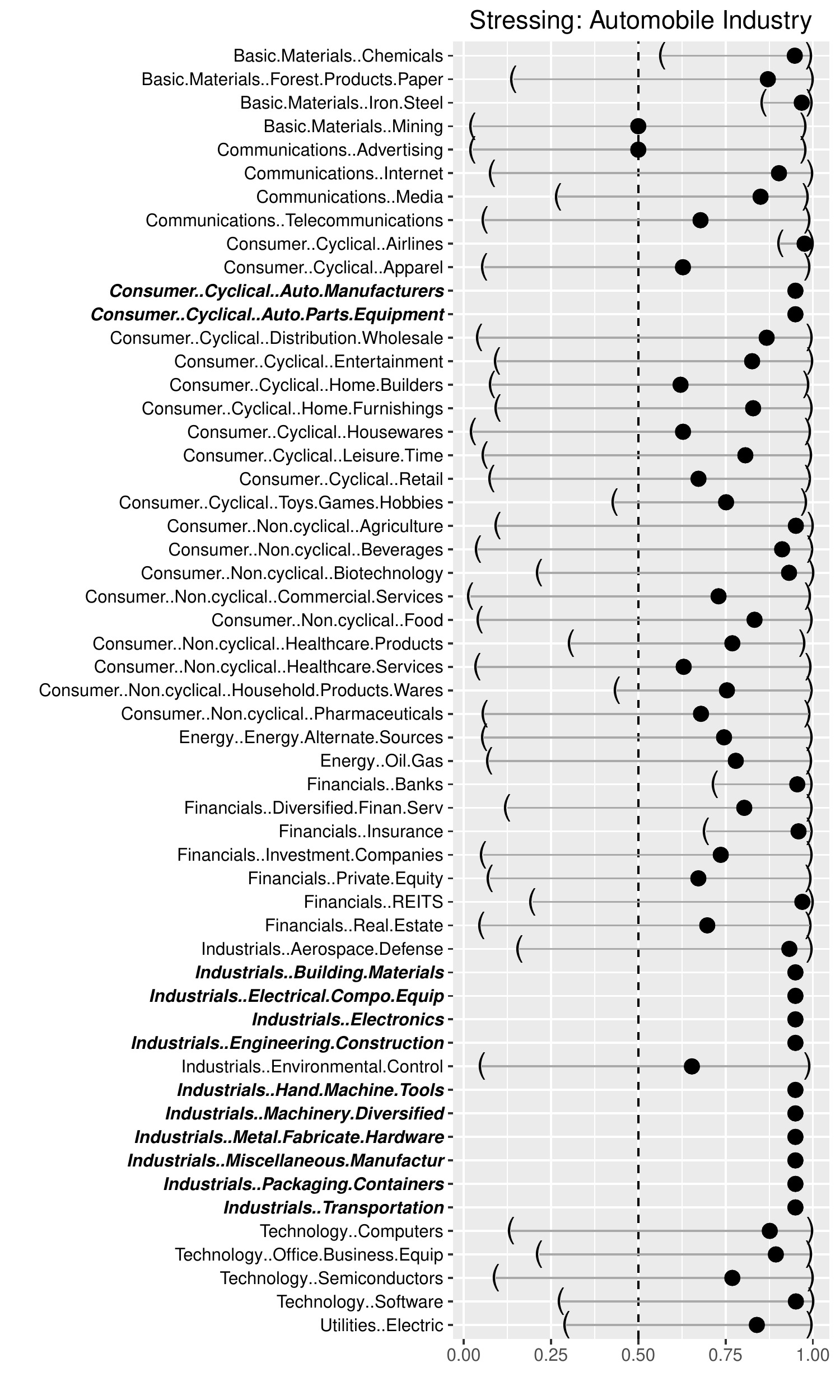}
	\caption{Results of stressing the sub-sectors Basic Materials (left) and the Automobile Industry (right). The bold italic sub-sectors are stressed at level 95\%.}
	\label{fig:stress_detailed_1}
\end{figure}

Generally, the Basic Materials sector is a category that accounts for companies involved with the discovery, development and processing of raw materials. The sector includes the mining and refining of metals (iron and steels), chemical producers and forestry products. As known from the theory, the Basic Materials sector is sensitive to changes in the business cycle. Similar to the proceeding in \ref{sec:aggr}, we do not vary the stress impulse within the sub-sectors of Basic Materials keeping it constant at 95\% and hence solely focus on the impact on other sub-sectors. In this case, a more granular insight is gained. For instance, the overall impact on Communications was observed on a medium level. On a more granular level and with reference to the left panel of \autoref{fig:stress_detailed_1}, the sub-sector Advertising turns out to be stress resistant.

Alternatively, the initial stress may start from a single sub-sector. This effect is exemplarily illustrated in the right panel of \autoref{fig:stress_detailed_1}, where we refined the previously conducted general scenarios of Cyclical Consumer Goods and Industrials of \ref{sec:aggr} by only inducing stress on the sector Automobile Industry represented by the sub-sectors Auto Manufacturers, Auto Parts \& Equipment, and several sub-sectors of the Industrials sector (see the sub-sectors written in bold and italics in the right panel of \autoref{fig:stress_detailed_1}). The Automobile Industry is probably one of the most important German industry sectors. In 2014, for instance, three of the four biggest German companies (ordered by sales volume) were producing automobiles (Volkswagen AG, Daimler and BMW). Against the current economic and political background, short-term or medium-term turmoils might result from the upcoming new American protectionism related to the discussions on the introduction of possible trade barriers imposed by the newly elected US government, but also if we consider the innovative processes and services related to electric mobility or the increasing interconnection between IT technology and car construction. As expected, we observe that this stress scenario has a severe impact on the entire German economy. 28 out of the remaining 43 sub-sectors exhibit a predicted conditional median greater than 75\%, 12 sub-sectors are strongly affected with conditional medians greater than 90\% and 7 (Iron and Steel, Airlines, Agriculture, Banks, Insurance, REITS and Software) even exceed the stress level of 95\%.

Despite looking at response industry sectors and covariates simultaneously, we also examined the impact of lagged covariates. However, these turned out to not have a relevant
influence on the response industry sectors, neither as single covariates nor in combination with non-lagged variables. This might be due to the fact that the sector PDs are derived from equity data and financial markets anticipate future depreciations in current stock prices.

\subsection{Results from alternative approaches}
\begin{figure}[htbp]
	\centering
	\includegraphics[width=0.69\linewidth]{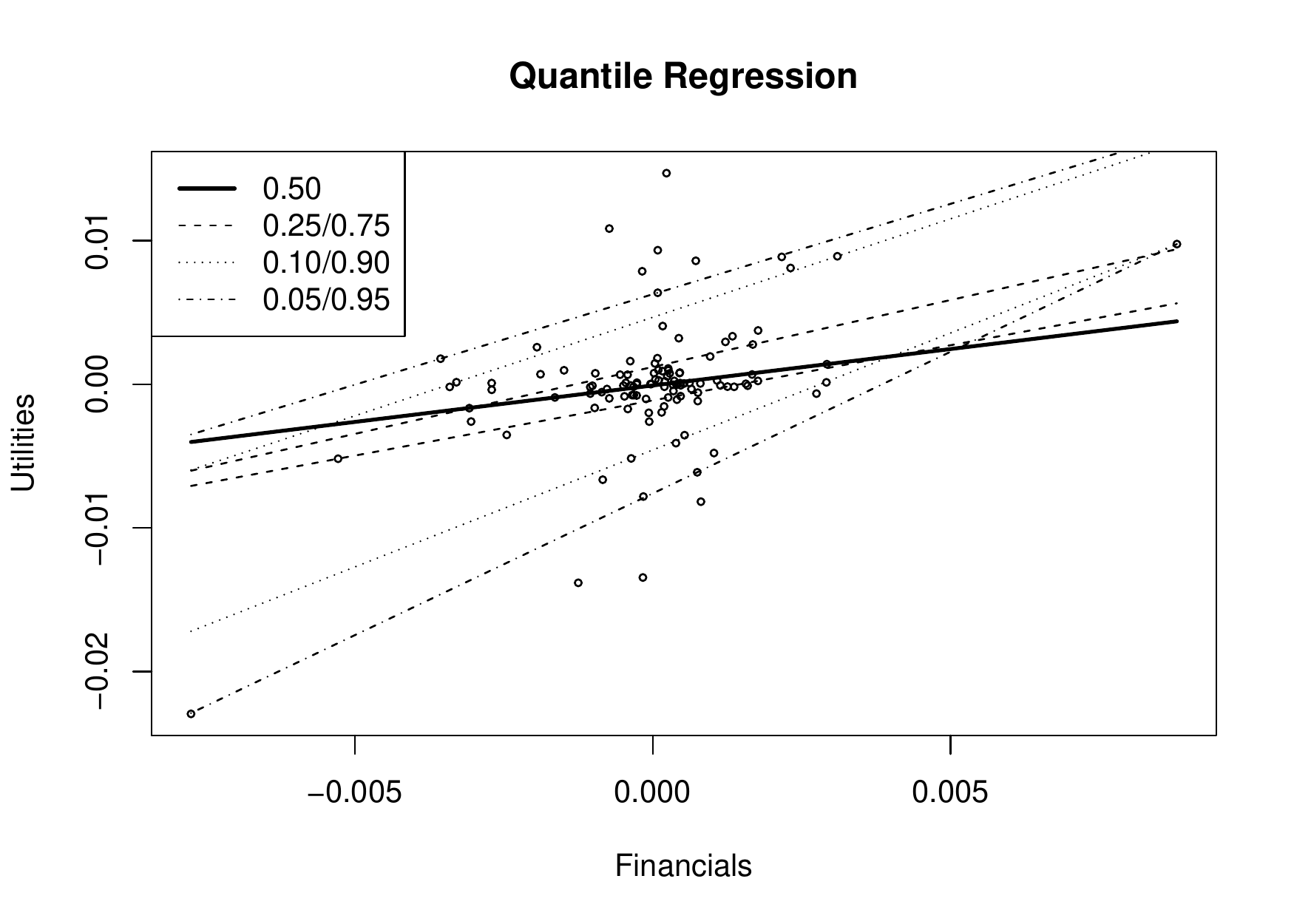}\\	\includegraphics[width=0.69\linewidth]{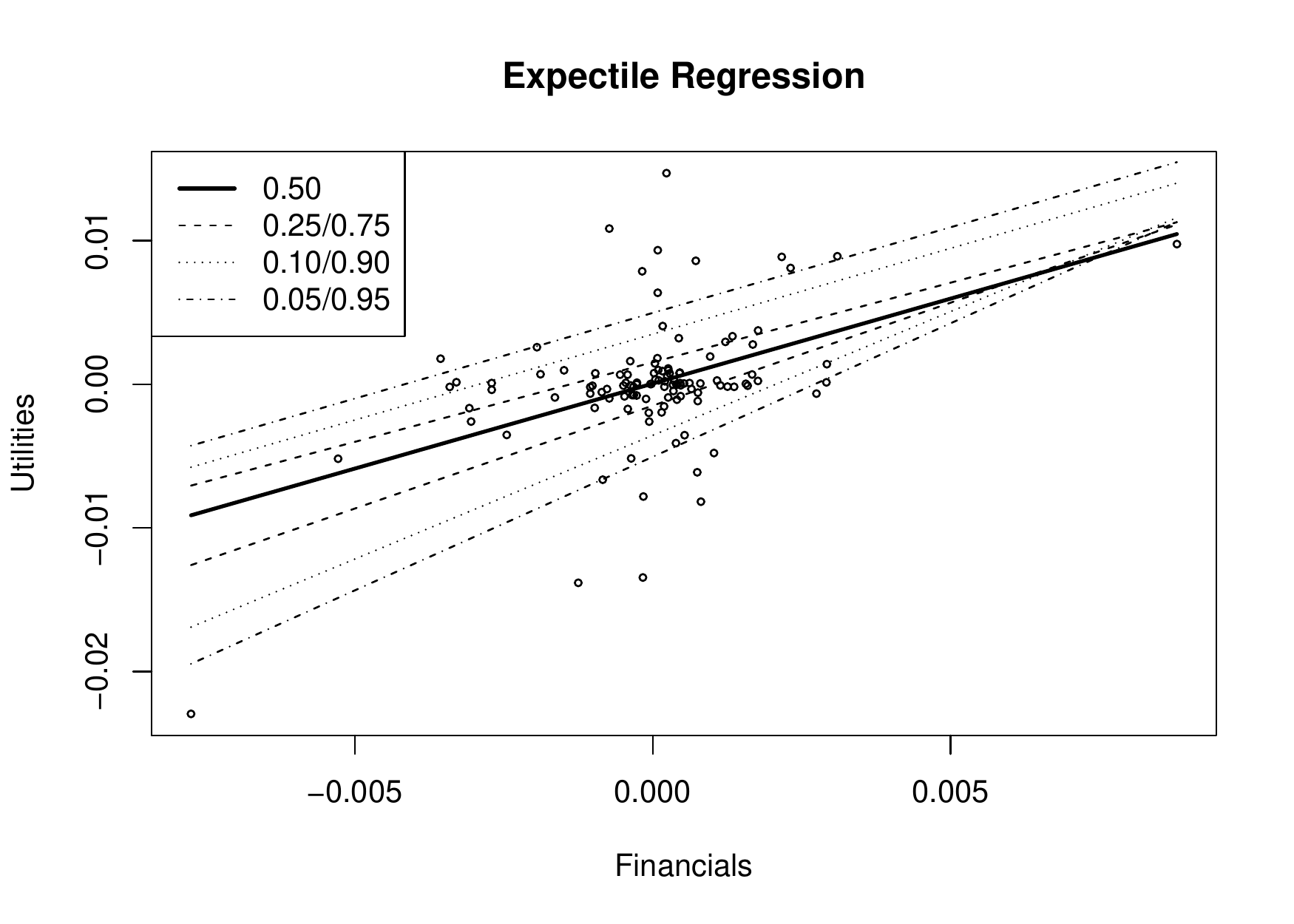}	\\\includegraphics[width=0.69\linewidth]{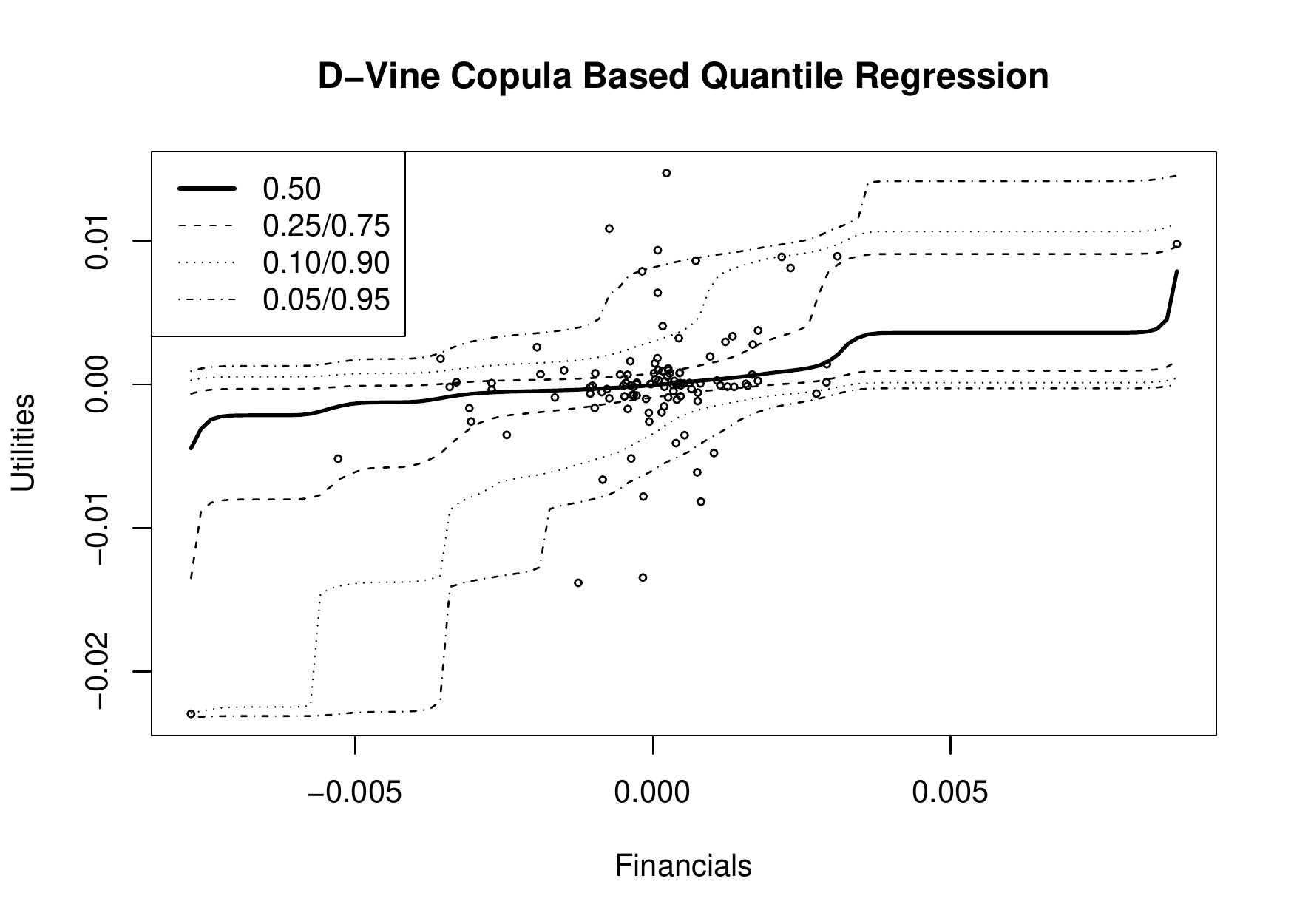}	
	\caption{Estimated curves of regressing the differenced PDs of Utilities against the different PDs of Financial using linear quantile regression (upper panel), expectile regression (middle panel) and D-vine copula based quantile regression (lower panel) for different levels of $\alpha$.} \label{fig:qc}
\end{figure}

In order to motivate the use of copula based quantile regression, we would like to point out some results from alternative approaches. Conditional quantiles
\[\arg \min_{q_{i,\alpha}}\sum_{i=1}^n ((1-\alpha)I(y_i<q_{i,\alpha})+\alpha I(y_i\geq q_{i,\alpha}))|y_i-q_{i,\alpha}|\]
can also be linearly modeled by traditional quantile regression, see e.g., \cite{koenker2006quantile} and for an example \autoref{fig:qc} where the method is applied to the differenced time series of response sector Utilities and the covariate Financials. However, as the conditional quantiles at different levels $\alpha_1$ and  $\alpha_2$ are estimated independently it can occur that for given covariate realizations and levels $\alpha_1<\alpha_2$, $q_{\alpha_1}$ can have a larger value than $q_{\alpha_2}$, an effect known as quantile crossing. Whereas different approaches have already been proposed to combat quantile crossing, see e.g., \cite{he1997quantile}, \cite{dette2008non} or \cite{bondell2010noncrossing}, the novel D-vine copula based quantile regression of \cite{kraus2017d} also ensures that this effect is prevented.\\

Expectile regression might be an alternative to quantile regression and has gained increasing attention in the recent literature, see e.g., \cite{waltrup2015expectile}. Based on replacing the L1 weighting scheme for conditional quantiles by an L2 metric, expectiles
\[\arg \min_{e_{i,\alpha}}\sum_{i=1}^n ((1-\alpha)I(y_i<e_{i,\alpha})+\alpha I(y_i\geq e_{i,\alpha}))(y_i-e_{i,\alpha})^2\]
can be bijectively linked to conditional quantiles and also uniquely determine a distribution function. Same as for quantile regression, expectile regression can lead to crossing of estimated curves, see the middle panel of \autoref{fig:qc}. Even though this problem is in practice less likely for expectile regression than for quantile regression, see e.g., \cite{schnabel2011expectile}, similar methods as for regularizing traditional quantile regression can be applied to prevent this effect. Although expectile regression implies some advantages such as computational efficiency, it still bears the problem of weak interpretability. Except for the $50$\% expectile which can be interpreted as a mean parameter, all other expectiles do not follow an intuitive interpretation in contrast to quantiles. As a consequence, modeling the $50$\% expectile is equivalent to applying a traditional linear model.

Due to the intuitive interpretability of quantiles and the robustness of the median, we favor quantile regression over expectile regression in this work. Moreover, following the criticism in \cite{bernard2015conditional} that conditional quantiles are only linear given very strong assumptions on the underlying copula of the response and covariate variables as well as the quantile crossing problem, we identify the D-vine copula based quantile regression approach as a suitable method for our stress test, see the lower panel of \autoref{fig:qc}.

\section{Summary and Outlook}\label{sec:conclusion}
This work evaluates the mutual impact of industry sector specific stress in the German economy using D-vine copula based quantile regression. Concerning simultaneous consideration of covariate and response variables, we illustrate the impact of sector stress arising from Basic Materials, Cyclical Consumer Goods or Industrials on the other industry segments. Above that, our analyses confirm that the Automobile sector strongly influences the whole economy. We also identify that a stressed Financials sector has only moderate influence on the other German industry. Similarly, lagged covariate segments exhibit only minor impact on their response sectors as well. We identify D-vine copula based quantile regression as a preferable method for conducting industry sector stress tests as it accounts for the nonlinearity properties of conditional quantiles and prevents quantile crossing effects. Moreover, in contrast to modeling conditional expectiles, conditional quantiles allow for an intuitive interpretation and have better robustness properties.


\vspace{6pt}

\bibliographystyle{apalike}
\bibliography{references}

\begin{thebibliography}{}

\bibitem[Aas et~al., 2009]{aasczado}
Aas, K., Czado, C., Frigessi, A., and Bakken, H. (2009).
\newblock Pair-copula constructions of multiple dependence.
\newblock {\em Insurance, Mathematics and Economics}, 44:182--198.

\bibitem[Bedford and Cooke, 2002]{bedford2002vines}
Bedford, T. and Cooke, R.~M. (2002).
\newblock Vines: A new graphical model for dependent random variables.
\newblock {\em Annals of Statistics}, 30:1031--1068.

\bibitem[Bernard and Czado, 2015]{bernard2015conditional}
Bernard, C. and Czado, C. (2015).
\newblock Conditional quantiles and tail dependence.
\newblock {\em Journal of Multivariate Analysis}, 138:104--126.

\bibitem[Bondell et~al., 2010]{bondell2010noncrossing}
Bondell, H.~D., Reich, B.~J., and Wang, H. (2010).
\newblock Noncrossing quantile regression curve estimation.
\newblock {\em Biometrika}, 97(4):825--838.

\bibitem[Covas et~al., 2014]{covas2014stress}
Covas, F.~B., Rump, B., and Zakraj{\v{s}}ek, E. (2014).
\newblock Stress-testing {US} bank holding companies: A dynamic panel quantile
  regression approach.
\newblock {\em International Journal of Forecasting}, 30(3):691--713.

\bibitem[Dette and Volgushev, 2008]{dette2008non}
Dette, H. and Volgushev, S. (2008).
\newblock Non-crossing non-parametric estimates of quantile curves.
\newblock {\em Journal of the Royal Statistical Society: Series B},
  70(3):609--627.

\bibitem[Fischer et~al., 2009]{fischeretal}
Fischer, M., K{\"o}ck, C., Schl{\"u}ter, S., and Weigert, F. (2009).
\newblock An empirical analysis of multivariate copula models.
\newblock {\em Quantitative Finance}, 9(7):839--854.

\bibitem[He, 1997]{he1997quantile}
He, X. (1997).
\newblock Quantile curves without crossing.
\newblock {\em The American Statistician}, 51(2):186--192.

\bibitem[Joe, 1997]{joe1997multivariate}
Joe, H. (1997).
\newblock {\em Multivariate models and multivariate dependence concepts}.
\newblock Boca Raton, FL: CRC Press.

\bibitem[Killiches et~al., 2017]{killiches2016examination}
Killiches, M., Kraus, D., and Czado, C. (2017).
\newblock Examination and visualisation of the simplifying assumption for vine
  copulas in three dimensions.
\newblock {\em Australian \& New Zealand Journal of Statistics}, 59(1):95--117.

\bibitem[Koenker, 2006]{koenker2006quantile}
Koenker, R. (2006).
\newblock Quantile regresssion.
\newblock {\em Encyclopedia of environmetrics}.

\bibitem[Koenker and Xiao, 2002]{koenker2002inference}
Koenker, R. and Xiao, Z. (2002).
\newblock Inference on the quantile regression process.
\newblock {\em Econometrica}, 70(4):1583--1612.

\bibitem[Kraus and Czado, 2017a]{kraus2017d}
Kraus, D. and Czado, C. (2017a).
\newblock D-vine copula based quantile regression.
\newblock {\em Computational Statistics and Data Analysis}, 110C:1--18.

\bibitem[Kraus and Czado, 2017b]{kraus2017growing}
Kraus, D. and Czado, C. (2017b).
\newblock Growing simplified vine copula trees: improving {D}i{\ss}mann's
  algorithm.
\newblock {\em arXiv preprint arXiv:1703.05203}.

\bibitem[Merton, 1974]{merton1974pricing}
Merton, R.~C. (1974).
\newblock On the pricing of corporate debt: The risk structure of interest
  rates.
\newblock {\em The Journal of finance}, 29(2):449--470.

\bibitem[Nelsen, 2006]{nelsen2006introduction}
Nelsen, R. (2006).
\newblock {\em An introduction to copulas, 2nd}.
\newblock New York, NY: Springer Science+Business Media.

\bibitem[Ong, 2014]{ong2014guide}
Ong, M. L.~L. (2014).
\newblock {\em A guide to IMF stress testing: methods and models}.
\newblock International Monetary Fund.

\bibitem[Schechtman and Gaglianone, 2012]{schechtman2012macro}
Schechtman, R. and Gaglianone, W.~P. (2012).
\newblock Macro stress testing of credit risk focused on the tails.
\newblock {\em Journal of Financial Stability}, 8(3):174--192.

\bibitem[Schnabel, 2011]{schnabel2011expectile}
Schnabel, S. (2011).
\newblock {\em Expectile smoothing: new perspectives on asymmetric least
  squares. An application to life expectancy}.
\newblock PhD thesis, Utrecht University.

\bibitem[Sklar, 1959]{Sklar}
Sklar, A. (1959).
\newblock Fonctions d\'{e} repartition \'{a} n dimensions et leurs marges.
\newblock {\em Publications de l'Instutut de Statistique de l'Universit\'{e} de
  Paris}, 8:229--231.

\bibitem[Spanhel and Kurz, 2015]{spanhel2015simplified}
Spanhel, F. and Kurz, M.~S. (2015).
\newblock Simplified vine copula models: Approximations based on the
  simplifying assumption.
\newblock {\em arXiv preprint arXiv:1510.06971}.

\bibitem[St\"ober et~al., 2013]{stoeber2013simplified}
St\"ober, J., Joe, H., and Czado, C. (2013).
\newblock Simplified pair copula constructions -- limitations and extensions.
\newblock {\em Journal of Multivariate Analysis}, 119:101--118.

\bibitem[Waltrup et~al., 2015]{waltrup2015expectile}
Waltrup, L.~S., Sobotka, F., Kneib, T., and Kauermann, G. (2015).
\newblock {Expectile and quantile regression - David and Goliath?}
\newblock {\em Statistical Modelling}, 15(5):433--456.

\end{thebibliography}


\end{document}